\documentclass[rnote]{aa}
\usepackage{epsfig}
\usepackage{graphics}

\newcommand{\be}{\begin{equation}}
\newcommand{\ee}{\end{equation}}

\begin{document}

\title{A graph of dark energy significance \\
 on different spatial and mass scales}

\author{P. Teerikorpi\inst{1}
  \and  P. Hein\"{a}m\"{a}ki\inst{1}
    \and  P. Nurmi\inst{1}
\and A.D. Chernin\inst{2}
\and M. Einasto\inst{3}
\and M. Valtonen\inst{1}
\and G. Byrd\inst{4}
}

\institute{Tuorla Observatory, Department of Physics and Astronomy, University of Turku, 21500 Piikki\"{o},
Finland
\and Sternberg Astronomical Institute, Moscow
University, Moscow, 119899, Russia
\and Tartu Observatory, 61602, T\~{o}ravere, Estonia
\and University of Alabama, Tuscaloosa, AL 35487-0324, USA
}

\authorrunning{P. Teerikorpi et al.}
\titlerunning{A graph of dark energy significance on different scales}

\date{Received / Accepted}

\abstract
{The current cosmological paradigm sees the formation and evolution
of the cosmic large-scale structure as governed
by the gravitational attraction of the Dark Matter (DM) and the repulsion
of the Dark Energy (DE).
}
{We characterize  the relative importance of uniform and constant dark energy, as given by the $\Lambda$ term in the standard $\Lambda$CDM cosmology,
in galaxy systems of different scales, from groups to superclusters. 
}
{An instructive "$\Lambda$ significance graph" is introduced where the matter-DE density ratio
$\langle \rho_{\mathrm{M}} \rangle / \rho_{\Lambda}$ for different
galaxy systems is plotted against the radius $R$. This
presents gravitation and DE dominated regions and shows
directly  the zero velocity radius, the zero-gravity radius, and the Einstein-Straus radius for
any fixed value of mass.
 }
{Example galaxy groups and clusters from the local universe illustrate the use of
the $\Lambda$ significance graph. These are generally
located deep in the gravity-dominated region 
$\langle \rho_{\mathrm{M}}\rangle / \rho_{\Lambda} > 2 $, being virialized.   Extended clusters and main
bodies of superclusters can reach down near the border line between gravity-dominated and DE dominated regions $\langle \rho_{\mathrm{M}}\rangle / \rho_{\Lambda} = 2$.
The scale--mass relation from the standard 2-point correlation function intersects this balance line near
the correlation lenght.  
}  
{The $\log \langle \rho_{\mathrm{M}} \rangle / \rho_{\Lambda} $ vs.
$\log R$ diagram is a useful and versatile way to characterize the dynamical state of systems of galaxies within the $\Lambda$ dominated expanding universe.
}


\maketitle

\section{Introduction}

According to the standard $\Lambda$CDM cosmology, a great majority of the material contents of the universe or about 95 per cent in energy density, is made of unknown dark substances called dark matter and dark energy (DE).

Dark matter is revealed by its gravitation, while dark energy (represented by Einstein's cosmological constant $\Lambda$
in standard cosmology) is an "antigravitating" uniform vacuum-like fluid.
The DE background produces antigravity. It  is stronger than
matter gravity in the present universe as a whole, making the
universal expansion accelerate.

According to the Planck Surveyor results (Planck Collaboration \cite{planck13}), the global density of the DE is, in round numbers,
$\rho_{\Lambda}
\approx 6 \times 10^{-30}$ g
cm$^{-3}$.\footnote{ The difference from the previous WMAP value ($\approx 7 \times 10^{-30}$ g
cm$^{-3}$) is mainly caused by the smaller value of the Hubble constant ($h = 0.67$) , lowering the
critical density ($1.88 h^2 \times 10^{-30}$ g
cm$^{-3}$) which is now equal to $\rho_{\mathrm{crit}} = 8.52 \times 10^{-30}$ g
cm$^{-3}$.}

Because of the non-uniform distribution of gravitating matter, the cosmic
antigravity can be stronger than gravity
also locally on scales of $\sim 1-10$ Mpc
(Chernin \cite{chernin01}),
which in principle makes it possible to detect the DE in the local galaxy universe, as reviewed by Byrd et al. (\cite{byrd12}, \cite{byrd15}).
For a recent study about the DE in the vicinity of the Local Group, see
Saarinen \& Teerikorpi (\cite{saarinen14}).

The presence of DE along with gravitating matter influences the formation of the large scale structure on all scales from groups of galaxies to superclusters. In regions where the DE dominates over the gravitating matter, new structures do not condense and linear perturbations of density decay (or even non-linear if sheet-like Chernin et al. \cite{chernin03}). The antigravity of DE also puts an upper limit on the size of a bound system of a certain mass and it influences dynamic mass determinations, leading to too low a mass Chernin et al. (\cite{chernin09}, \cite{chernin12}).

It occurs that a useful parameter which characterizes the influence of DE is the energy density ratio
$\langle \rho_{\mathrm{M}} \rangle / \rho_{\Lambda}$ as calculated for
the system or scale under inspection.

In the present Note, we introduce a graph presenting the ratio $\langle \rho_{\mathrm{M}} \rangle / \rho_{\Lambda}$  for systems of different size and mass.
This diagram displays in a direct way a few relevant scales which appear for each fixed mass of a spherical or slightly flattened system in $\Lambda$CDM cosmology. The location of a galaxy system in the diagram indicates whether
its overall dynamics is dominated by gravity or DE antigravity.
This $\Lambda$ significance graph has also limited use for flattened systems. 

\section{Zero-gravity radius and Einstein-Straus scale} 

 A natural scale which appears around a mass point (or a spherically symmetric system) embedded in
the dark energy background, is the zero-gravity radius $R_{\rm ZG}$.
 In a weak field situation
one may write (Chernin et al. \cite{chernin09})
the force affecting a test particle with mass $m$ as
the sum of Newton's gravity force
produced by the mass $M$ and 
Einstein's antigravity force due to DE:
\be F(R) = \big (- \frac{GM}{R^2} + \frac{8\pi G}{3}
\rho_{\Lambda} R \big) m. \ee

At the zero-gravity distance $R = R_{\rm ZG}$  (Chernin \cite{chernin01}): 
\be R_{\rm ZG} = \big (\frac{M}{\frac{8\pi}{3} \rho_{\Lambda}}\big )^{1/3}
= 1.1\, \mathrm{Mpc} \times \big (\frac{M / 10^{12} M_{\odot}}{\rho_{\Lambda}/6 \times 10^{-30} \mathrm{g/cm}^3}\big )^{1/3}
 \ee
\noindent
gravity is equal to antigravity.

Another interesting scale is related to the Einstein-Straus radius $R_{\mathrm{ ES}}$. This comes from the central mass plus vacuole solution by Einstein \& Straus
(\cite{einstein45}), where the radius of the vacuole is such that the mean density within it is equal to
the current density of the Friedmann universe:

\be R_{\mathrm{ ES}} = [M /(\frac{4 \pi}{3} \rho_{\rm m})]^{1/3} = R_{\mathrm{ZG}}
(2\frac{\rho_{\Lambda}}{\rho_{\rm m}})^{1/3}
 \simeq 1.7 R_{\mathrm{ZG}}
\ee 
This may be imagined as the radius of the spherical volume from which the mass making
up the system has been gathered in the past during the gravitational instability process. 

Dark energy
puts an absolute upper limit on the size of a gravitationally bound system: it
must be located within its zero-gravity
sphere where gravity dominates. 

\section{Matter-to-DE ratio
 vs. scale diagram}

It is instructive and useful to construct a diagram where the $x$-axis is the logarithm of the spatial scale (or the
radius of a system) and the $y$-axis gives the (log) ratio of the average density $\langle \rho_{\mathrm{M}}\rangle$ of a system and the (constant) DE density equal to the global value $\rho_{\Lambda}$.

For $\rho_{\Lambda}
\approx 6 \times 10^{-30}$ g
cm$^{-3}$, the ratio $\langle \rho_{\mathrm{M}}\rangle / \rho_{\Lambda}$ becomes
\be
\langle \rho_{\mathrm{M}}\rangle / \rho_{\Lambda} = 2.7 \frac{(M/10^{12}M_\odot)}{(R/\mathrm{Mpc})^3}
\ee
where $M$ is the mass within the radius  $R$.
In logarithms: 
\be
\log \langle \rho_{\mathrm{M}}\rangle / \rho_{\Lambda} = 0.43 + \log M/10^{12}M_\odot
- 3 \times \log R/\mathrm{Mpc}
\ee
In the $\log \langle \rho_{\mathrm{M}}\rangle / \rho_{\Lambda}$ vs. $\log R$ diagram
this relation forms a family of inclined straight lines for each fixed value of 
the mass $M$. Fig. 1 explains the meaning of the different lines and regions
in such a diagram. Note that the fraction of the DE density grows when one goes downwards in the diagram.

In the upper part of the diagram, the systems are dominated by gravitation. 
They may be bound and
for sufficiently high $\langle \rho_{\mathrm{M}}\rangle / \rho_{\Lambda}$
they can be virialized (Sect.5).  

We show in Fig.1 the horizontal line giving the zero-velocity radius
$R_{\mathrm{ZV}}$ when
intersecting the inclined constant mass line. This density ratio is
obtained from the classical relation between $M$ and $R_{\mathrm{ZV}}$
\be
M = \frac{\pi^2}{8G}t_0^{-2}R_{\mathrm{ZV}}^3 = 2.74 \times 10^{12} M_\odot (\frac{t_0}{10^{10}\mathrm{yrs}})^{-2}(\frac{R_{\mathrm{ZV}}}{1 \mathrm{Mpc}})^3
\ee 
by Lynden-Bell (\cite{lyndenbell81}). In addition, the effect of $\Lambda$ 
increases $R_{\mathrm{ZV}}$ by about 1.15, or the inferred mass by about 
1.5 (Peirani \& de Freitas Pacheco \cite{peirani06}; Saarinen \& Teerikorpi \cite{saarinen14}; 
Tully \cite{tully15}). The calculation leads to  
$\log \frac{\langle \rho_{\mathrm{M}} \rangle}{\rho_{\Lambda}} \approx 0.75$.

The lower part of the diagram contains two horizontal lines. One is for the
ratio  $\langle \rho_{\mathrm{M}}\rangle / \rho_{\Lambda} = 2$
(gravity = antigravity). The second one is for the global
matter-to-DE ratio $0.43$. Spherical systems below the upper line are dynamically dominated by the antigravitating DE background.

When we consider systems of a fixed mass $M$ and go down
along the inclined straight line (Eq.(5)), it  
intersects the upper line at the zero-gravity radius $R_{\rm ZG} (M)$
(Eq.(2)), which also is the maximal possible size of a bound system. 

For systems below the upper horizontal  line the radius of the system exceeds the zero-gravity radius. Then the system is not 
gravitationally bound as a whole. Note, however, that this is valid for
a spherical system.

Furthermore, the point of intersection with the "global" line gives formally  
the Einstein-Straus radius $R_{\mathrm{ ES}}$ for this same central mass (Eq.(3)). This is also the distance where the Hubble flow around
the mass reaches the global Hubble expansion rate (Teerikorpi \& Chernin \cite{teerikorpi10}).
Below the global line, there is matter underdensity.

In a perfectly
uniform Friedmann model, only the points of the global line exist. Then $R$ refers to the current size of a comoving volume containing the matter mass $M$. 
\footnote{If we calculate $R_{\rm ZG}$ around a point
in a homogenous world from Eq.(2), this radius will increase directly
proportional to the radius of the considered matter sphere. This means that
one cannot ascribe physical significance to the zero-gravity radius within a fully uniform universe -- every point is as it were on the surfaces of a great
number of zero-gravity spheres of arbitrarily different sizes, not feeling any force.}
 
In Fig.1 we have also shown what happens if the
mass of a system is $M_1$ as measured within a radius $R_1$,
and then increases (e.g., $M \propto R$),
up to a zero-gravity radius $R_{\rm ZG}'$,
the absolute upper limit for the size of a bound system.
Chernin et al. (\cite{chernin12}) considered such a situation for
galaxy groups using different mass-radius laws (density profiles).

\begin{figure}
\epsfig{file=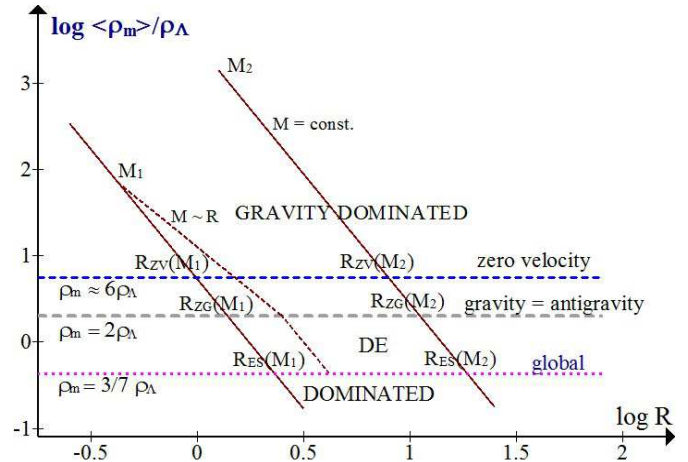 , angle=0, width=8.7cm}
\caption{$\log \langle \rho_{\mathrm{M}} \rangle / \rho_{\Lambda} $ vs.
$\log R$ for spherical systems. The inclined lines
correspond to different mass values. Above the middle "gravity = antigravity" line the region is dynamically 
dominated by gravitation, below this line by DE. Intersections give the radii
$R_{\mathrm{ZV}}$,
 $R_{\mathrm{ZG}}$, and $R_{\mathrm{ES}}$.
Dotted inclined lines illustrate the case when the mass increases with the radius (see the text).
}
\end{figure}

\section{Example systems of galaxies}

To illustrate the meaning of the  $\langle \rho_{\mathrm{M}} \rangle / \rho_{\Lambda} $ vs. scale diagram, we calculate for several well-known galaxy systems (listed below)
the ratio $\langle \rho_{\mathrm{M}} \rangle / \rho_{\Lambda} $ and put
them in the $\langle \rho_{\mathrm{M}} \rangle / \rho_{\Lambda} $ vs. $\log R$
diagram of Fig.2.   

In some
cases we plot the results for two or more increasingly large radii around the system. The mass and size values are characteristic rather than accurate.

{\it The Local Group}.
The mass of the LG has notable uncertainty, with values cited
from $1.5$ to about $4 \times 10^{12} M_\odot$ (e.g., van den Bergh 
\cite{vandenbergh99}; Chernin et al. \cite{chernin09}; Partridge et al.
\cite{partridge13}). Here we use the mass value $2 \times 10^{12} M_\odot$
and the radius $ R =1.0$ Mpc. Also, we have plotted the zero-gravity region, $R_{\rm ZG} = 1.4$ Mpc, in order to illustrate its location at the
intersection of the two lines as noted above.

{\it The Local Sheet}.
This local flattened system comprises, in addition to the Local Group, other
groups and galaxies within about 6 Mpc and seems to form a dynamically separate
system, not being just an arbitrarily defined part of the plane of the Local Supercluster. McCall (\cite{mccall14}) estimates its mass as about 2.7 times that of
the LG, or in our diagram $5.4 \times 10^{12} M_\odot$, within the radius
of 3.75 Mpc.

{\it The Fornax Cluster}.
For an inner part of this cluster ("Fornax$_1$" in our diagram) we use the virial mass
 $7 \times 10^{13} M_\odot$ (Drinkwater et al. \cite{drinkwater01}) for $R_{\mathrm{vir}} = 1.4$ Mpc. 
Using a Tolman-Bondi infall model (with $\Lambda$), Nasonova et al. 
(\cite{nasonova11})
derived $2.2 \times 10^{14} M_\odot$ within $4.6$ Mpc
("Fornax$_2$").

{\it Virgo and the Local Supercluster}. For the central part
of the LSC, the Virgo cluster ("Virgo$_1$"), we  use its
virial mass $\sim 1 \times 10^{15} M_\odot$ within $2$ Mpc
(e.g., Teerikorpi et al. \cite{teerikorpi92}; Ekholm et al. \cite{ekholm00}). 
Within the LG distance
(18 Mpc) ("Virgo$_2$"), the mass is about
$5 \times 10^{15} M_\odot$, from a TB
analysis (Ekholm et al. \cite{ekholm00}).

{\it The Coma Cluster}. Chernin et al. (\cite{chernin13}) analyzed the observational
results on the mass of the Coma cluster at different distances from the
centre using different mass profiles and considerig the influence
of DE. They concluded that at 1.4 Mpc, the mass is $4.4 \times 10^{14} M_\odot$ (Coma$_1$), at 4.8 Mpc,
$2.6 \times 10^{15} M_\odot$ (Coma$_2$), and at 14 Mpc, $5.4 \times 10^{15} M_\odot$ (Coma$_3$). The mass value at 14 Mpc takes into account DE.

{\it The Perseus-Pisces Supercluster}. Hanski et al. (\cite{hanski01}) derived
for the inner part of the PP supercluster the virial mass value of about $1.5 \times 10^{15} M_\odot$ within  2.6 Mpc (PP$_1$). For the larger region within
22 Mpc (PP$_2$) the value $8 \times 10^{15} M_\odot$ was obtained from an analysis based on the Tolman-Bondi model including the $\Lambda$ term.

{\it The Corona Borealis supercluster}. 
Pearson et al. (\cite{pearson14})
found that the central parts of the Corona Borealis supercluster (CB) could have
a wide mass range of about $0.6 - 12 \times 10^{16} h^{-1} M_\odot$
within a radius of $12.5 h^{-1}$Mpc, the likely value being around
$1 \times 10^{16} h^{-1} M_\odot$ . We use the representative
values $1.5 \times 10^{16} M_\odot$ within $19$ Mpc.

{\it The Laniakea Supercluster}. This supercluster (Tully et al. 2014) is the largest well-studied aggregation of matter in the observable universe. 
According to Tully et al. (\cite{tully14}), the supercluster, if approximated as round, has a radius
of 80 Mpc and a mass of about $10^{17} M_\odot$ which comes from the volume and the  mean cosmic density. No independent total mass estimate is available.

Also, the region of {\it dwarf galaxy associations} (DGAs) is shown in the
$\log \langle \rho_{\mathrm{M}} \rangle / \rho_{\Lambda} $ vs.
$\log R$ diagram of Fig.2.
They have radii $\sim 0.3$ Mpc and virial masses within $(0.05 -1) \times 10^{12} M_\odot$ (refs. in Chernin \& Teerikorpi
\cite{chernin14}). 
We have also marked the location ({\it SDSS groups}) around which the galaxy groups (with less than 50 members) extracted from the SDSS-DR10 survey are concentrated
(masses from the virial theorem; Tempel et al. \cite{tempel14}). 

\begin{figure}
\epsfig{file=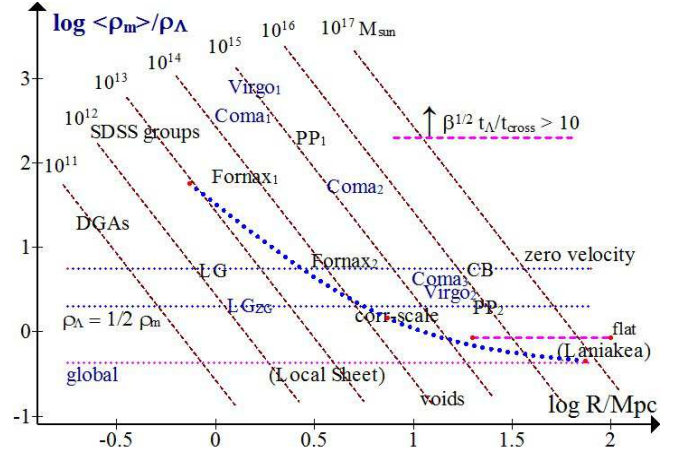 , angle=0, width=8.5cm}
\caption{$\log \langle \rho_{\mathrm{M}} \rangle / \rho_{\Lambda} $ vs.
$\log R/\mathrm{Mpc}$ for example galaxy systems. The inclined lines
are for the mass values indicated in $M_\odot$. The CB supercluster has been
slightly shifted to the right. The dotted curve indicates the characteristic mass within radius
$R$ as given by the standard 2-point correlation function. The "flat" balance line
refers to the limiting case of a flattened system.
}
\end{figure}

\section{Discussion}

The individual objects in Figure 2 represent a heterogeneous collection of examples
from the nearby universe. Nevertheless it is
interesting to note that the systems are generally found above
the "gravity = antigravity" $\langle \rho_{\mathrm{M}} \rangle = 2  \rho_{\Lambda} $ line, which defines the absolute upper size
for a gravitationally bound spherical system. 

Prominent groups and clusters
are usually located deep within the gravity dominated radius ($R$
much less than $R_{\mathrm{ZG}}$), which is also a requirement
for them to be virialized. The virial theorem in general form (Chernin et al. \cite{chernin09}) is

\be
M = \beta \sigma_V^2 R/G + \frac{8\pi}{3}\rho_{\Lambda} R^3
\ee
where $\beta$ is a constant $\sim $ unity. Dividing by $(4\pi /3)R^3
\rho_{\Lambda}$ and inserting the timescale
 $t_{\Lambda} = (((8\pi/3)G\rho_{\Lambda} )^{-1/2}$ one obtains

\be
\frac{\langle \rho_{\mathrm{M}} \rangle}{\rho_{\Lambda}} =
2(1+\beta ( \frac{t_\Lambda}{t_{\mathrm{cross}}})^2)
\ee
%
where the crossing time $t_{\mathrm{cross}} = R/ \sigma_V$ should be considerably smaller than $t_\Lambda \approx 17 \times 10^9 \mathrm{yrs}$ for a virialized system. 
We indicate in Fig.2 the region where $\beta^{1/2} \frac{t_{\Lambda}}{t_{\mathrm{cross}}}>10$. There also $R$ is $\ga 5$ times smaller than $R_{\mathrm{ZG}}$. 

We note that going upwards along a constant-$M$ line makes the influence of the DE
on the virial mass determination decrease. It can be shown that the mass derived
from Eq.(7) is a factor of $(1+\alpha (R/R_{\mathrm{ZG}})^3)$ larger
than the classical result (here $\alpha \sim 1$). The increase is insignificant for
well-virialized systems.

In Fig.2 the systems above the "zero-velocity line"
(Sect.3) are totally within
the collapsing region. The outer parts of systems below the line have not
been retarded down to zero velocity (e.g., the extended Fornax$_2$, Coma$_3$,
and Virgo$_2$ systems).

For example, we can read from the diagram that if the mass of the Virgo cluster only would control the dynamics
of the Local supercluster up to the distance of the LG,
then the zero-gravity distance (at around 11 Mpc) would lie closer than the LG
and we would be in the DE dominated region. However,
the mass between us and Virgo seems to
keep us near the gravity dominated region (Virgo$_2$ in Fig.2). 

The main body of the Corona Borealis supercluster lies in Fig.2 near the zero-velocity
line.  According to Pearson et al. (\cite{pearson14}), this is an
example when a supercluster made of several rich clusters may be bound,
containing much intercluster matter.

The Local Sheet and the Laniakea supercluster are found below the
$\langle \rho_{\mathrm{M}} \rangle = 2 \rho_{\Lambda} $ line. However,
the Local Sheet is a strongly flattened system (hence in brackets) and the condition
$\langle \rho_{\mathrm{M}} \rangle = 2 \rho_{\Lambda} $ for "gravity = antigravity"
at the outer border is not valid.

A flattened system, if viewed as a disc, cannot  be bound as a whole if the antigravity
force between its centre and the edge (one radius away) is larger than the gravity
pulling the edge inward. It is known
that a disc exerts a larger gravity force on a particle at its edge than a sphere  of the same mass and radius. If the ratio of these
forces is $\alpha (>1)$, then Eq.1 leads to the balance condition
$\langle \rho_{\mathrm{M}} \rangle = (2/\alpha) \rho_{\Lambda}$. Therefore
the "gravity = antigravity" line should be lowered for a flattened system.
Here $\langle \rho_{\mathrm{M}} \rangle $ is the mean density
caused by the mass $M$ within the sphere having the radius $R$
(not the mean density of the disc).

For a flat disc, with the density increasing inwards, the ratio $\alpha < 3\pi/4$, while
the limiting case of constant density gives $\alpha = 3\pi/4 = 2.35$ (see e.g., Woltjer 1967). In this limiting case, the $x$ coordinate of the balance line in Fig. 2 would be shifted down by 0.37 so that $\langle \rho_{\mathrm{M}} \rangle / \rho_{\Lambda}
= -0.07$
in order to roughly take into account flattened systems.

Chernin et al. (\cite{chernin15}) conclude that the
Local Sheet is
expanding with acceleration and has under-density. This would agree
with its location even below the global density line in Fig.2.
In fact, the zero-gravity spheres around the LG and nearby
groups do not appear to intersect, also suggesting accelerating
Hubble recession (Byrd et al. \cite{byrd12}).

The location of the Laniakea supercluster near the "global" line simply reflects
the way its mass was estimated from the mean cosmic density by Tully et al. (\cite{tully14}).
Even without an independent mass estimate it is clear that the supercluster as a prominent mass concentration
hardly can lie below the global line (in the region of voids).
On the other hand, if it were to be found near or above the $\langle \rho_{\mathrm{M}} \rangle = 2 \rho_{\Lambda}$ line, its mass had to be the very large $\ga 4 \times 10^{17} M_\odot$, while
the 
large-scale structure formation theory suggests an upper limit of about $
10^{16} M_{\odot}$ for the massive {\it bound}
objects (e.g., Holz \& Perlmutter \cite{holz12}).

The $\Lambda$ significance diagram may also throw light on the general galaxy field as described by the
two-point correlation function (without biasing).
With the values generally regarded as standard, i.e. the correlation length $r_0 \approx
5 h^{-1}_{100}$ Mpc and the correlation exponent $\gamma \approx 1.75$, we
have calculated, as explained in Teerikorpi et al. (\cite{teerikorpi05}),
typical mass values within the radius $R$.
These have been plotted, as a dashed curve, on the diagram of Fig.2. As already noted by Teerikorpi et al. (\cite{teerikorpi05}), the correlation length $r_0$ is not far from
the zero-gravity radius corresponding to the fluctuation described by the
correlation function. Beyond $r_0$ the decreasing
fluctuations lie in the DE dominated region, where also the pairwise velocity
dispersion of the Hubble flow starts to diminish.

\section{Concluding remarks}

The $\log \langle \rho_{\mathrm{M}} \rangle / \rho_{\Lambda} $ vs.
$\log R$ diagram is a useful way to characterize systems of galaxies.
Different regions in the diagram correspond to the mass and size  of a system and its
dynamical state within the $\Lambda$ dominated expanding universe.

The $\Lambda$ significance graph will be used in forthcoming separate studies
to discuss clusters and superclusters, especially as extracted
from the SDSS survey. A preliminary inspection of 
the distribution of SDSS DR7 superclusters
shows a range of $\langle \rho_{\mathrm{M}} \rangle / \rho_{\Lambda}$ and an interesting dependence on the size and morphology of the system, which also can be studied using simulated systems. 

\section*{Acknowledgements}
We thank G. Bisnovatyi-Kogan, L. Liivam\"{a}gi and E. Saar for useful discussions. 
ME was supported by the ETAG project
IUT26-2 and by the European Structural Funds
grant for the Centre of Excellence "Dark Matter in (Astro)particle Physics
and Cosmology" TK120.

{}

\end{document}